\input{aipcheck}

\documentclass[
    ,final         ]
  {aipproc}

\layoutstyle{6x9}

\begin{document}

\title{The $^{85}$Rb(p,n)$^{85}$Sr reaction and the modified proton optical potential}


\author{G. G. Kiss}{
  address={Institute of Nuclear Research (ATOMKI), H-4001 Debrecen, POB. 51, Hungary}
}

\author{Gy. Gyürky}{
  address={Institute of Nuclear Research (ATOMKI), H-4001 Debrecen, POB. 51, Hungary}
}

\author{A. Simon}{
  address={Institute of Nuclear Research (ATOMKI), H-4001 Debrecen, POB. 51, Hungary}
}  

\author{Zs. Fülöp}{
  address={Institute of Nuclear Research (ATOMKI), H-4001 Debrecen, POB. 51, Hungary}
}  

\author{E.~Somorjai}{
  address={Institute of Nuclear Research (ATOMKI), H-4001 Debrecen, POB. 51, Hungary}
}  

\author{T.~Rauscher}{
  address={Universit\"at Basel, CH-4056 Basel, Switzerland}
}  

\begin{abstract}
The cross sections of the astrophysically relevant $^{85}$Rb(p,n)$^{85}$Sr$^{g,m}$
reaction have been measured
between E$_{c.m.}$ = 2.16 and 3.96 MeV. The cross sections have been derived by
measuring the $\gamma$ radiation following the $\beta$ decay of the reaction
products. A comparison with
the predictions of Hauser-Feshbach calculations using the
NON-SMOKER code confirms a recently derived modification of the global optical proton potential.
\end{abstract}

\maketitle

\section{Introduction}

The synthesis of the so-called $p$ nuclei \cite{woo78, arn03} (the heavy, proton-rich isotopes which cannot be synthesized by
neutron capture reactions in the s- or r-process) is still one of the least known processes of
nucleosynthesis. It is generally accepted that the synthesis of the $p$ nuclei, the astrophysical
p-process, mainly involves $\gamma$-induced reactions on abundant seed nuclei produced at earlier
stages of nucleosynthesis by the s- or r- process. 
During the p-process, material from the bottom of the valley of stability is driven to the proton-rich side by
consecutive ($\gamma$,n) reactions. As the neutron separation energy increases while the charged particle
separation energies decrease along this path, charged-particle emitting ($\gamma,\alpha$) and
($\gamma$,p) reactions become increasingly important for the more proton-rich region. 

Theoretical investigations show that in the case of the
production of the light $p$ nuclei, ($\gamma$,p) reactions play a key role \cite{rau06,rap06}. The relevant
astrophysical reaction rates can be determined from the cross
section of the inverse capture reactions through the detailed balance theorem if the corresponding capture cross sections are known experimentally. In the last few years several proton capture cross section measurements for $p$ process studies were carried out \cite{lai87,sau97,bor98,chl99,gyu01,har01,ozk02,gal03,gyu03,tsa04,gyu06, gyu07, kiss07, kiss08, fam08, spy08}.
However, it has been shown recently that not only ($\gamma$,p)
reactions are important for modeling the synthesis of the light $p$ nuclei, but (n,p) and (p,n) reactions should also be taken into account \cite{rap06}.

To investigate the impact of nuclear reaction rates on predicted $p$ process abundances, simulations with different sets of neutron, proton, $\alpha$-capture and photodisintegration rates have been performed by Rapp \textsl{et al}., \cite{rap06}. It is stated that some (p,n) reactions --- such as the $^{85}$Rb(p,n)$^{85}$Sr --- exhibit strong influence on the final $p$ abundances. 

Contrary to the well studied (p,$\gamma$) reactions, there is only limited experimental information available about the low-energy (p,n) cross sections in the mass region. Recently, the cross section of the $^{76}$Ge(p,n)$^{76}$As reaction has been measured \cite{kiss07} and considerable discrepancies between the experimental data and the theoretical prediction were found. To reproduce the cross sections the strength of the imaginary part of the widely-used semi microscopic potential of \cite{jeu77} (including low-energy modifications by \cite{lej80}) had to be increased by approximately 70\%. 

The above considerations show that certain (p,n) reactions influence directly the results of a $p$ process network calculations and they also provide a sensitive probe for the statistical model calculations. Therefore it is advised to continue the systematic study of nuclear reactions relevant for the $p$ process by measuring the cross section of the $^{85}$Rb(p,n)$^{85}$Sr. 

\section{Experimental technique}

The experiment was similar to our previous $^{76}$Ge(p,n)$^{76}$As studies \cite{kiss07}. Here a brief summary on the experimental details is given.

\begin{table}
\caption{\label{tab:decay}Decay parameters of $^{85}$Rb(p,n)$^{85}$Sr reaction
products taken from \cite{NDS}.}
\begin{tabular}{cccc}
\parbox[t]{0.6cm}{\centering{Residual \\ nucleus}} &
\parbox[t]{0.6cm}{\centering{Half-life}} &
\parbox[t]{2.8cm}{\centering{Gamma energy [keV]}} &
\parbox[t]{2.8cm}{\centering{Relative $\gamma$-intensity \\ per decay [\%]}} \\
\hline\hline
$^{85}$Sr$^g$ &  64.84 $\pm$ 0.02 d & 514.01 $\pm$ 0.02 & 96 $\pm$ 4 \\
$^{85}$Sr$^m$ & 67.63 $\pm$ 0.04 m & 231.64 $\pm$ 0.01 & 84.4 $\pm$ 0.2 \\
\hline
\end{tabular}

\end{table}

\begin{figure}
  \rotatebox{270} {\includegraphics[height=.58\textheight]{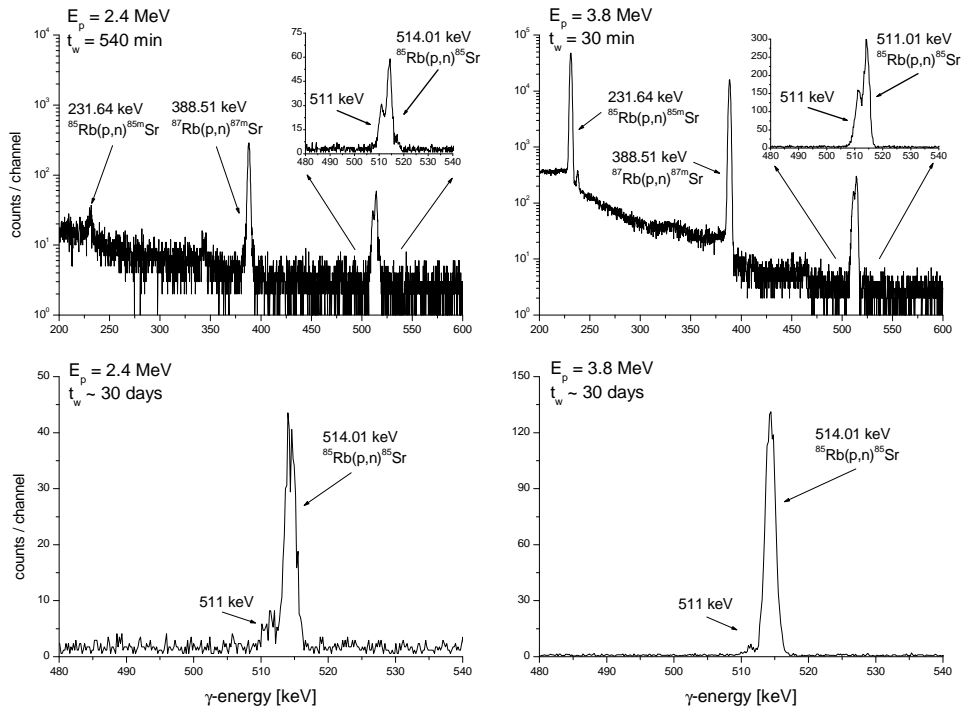}}
  \caption{Typical activation $\gamma$-spectra taken after the irradiation of RbCl target with 2.4 (left panel) and 3.8 MeV (right panel) proton beam. The 514 keV peak from the $^{85}$Rb(p,n)$^{85}$Sr$^g$ reaction can be well separated from the annihilation peak as can be seen on the insets. The length of the waiting time (t$_w$) between the end of the irradiation and the start of the $\gamma$-countings were 540 (E$_p$ = 2.4 MeV) and 30 min (E$_p$ = 3.8 MeV). The lower panels shows typical spectra taken in the repeated activity measurement approximately one month after the irradiations (for details see the text).}
\end{figure}

The targets were produced with evaporating natural RbCl (chemical purity: 99.99\%) onto thin Al foil. The isotopic abundances of $^{85}$Rb and $^{87}$Rb are 72.17 and 27.83\% respectively \cite{NDS}. The absolute number of target atoms and the uniformity were determined by Rutherford Backscattering Method (RBS) using the Nuclear Microbeam facility of ATOMKI \cite{simon06}. The energy of the proton beam provided by the Van de Graaff and cyclotron accelerators of ATOMKI was between 2 and 4 MeV (covered with 200 keV steps) with 600 nA beam current. Each irradiation lasted approximately 8 hours. The check the possible systematic errors, the E$_p$ = 2.6 MeV irradiation was carried out with both the Van de Graaff and the cyclotron accelerators and no difference in the cross section was found.

The $^{85}$Rb(p,n) reaction leads to ground ($^{85}$Sr$^g$) and isomeric states $^{85}$Sr$^m$ of the Strontium isotope.
The $^{85}$Sr$^g$ decays by $\beta$$^+$ to $^{85}$Rb and the $^{85}$Sr$^m$ with internal transition to the ground state of $^{85}$Sr and with electron capture and $\beta$$^+$ to $^{85}$Rb. For determining the cross section of the $^{85}$Rb(p,n)$^{85}$Sr$^g$ reaction the 514.01 keV, for the $^{85}$Rb(p,n)$^{85}$Sr$^m$ reaction the 231.84 keV gamma line was used. The decay parameters of $^{85}$Sr$^{g,m}$ isotopes are summarized in Table I. Since proton induced reactions on RbCl are leading to stable or short lived isotopes - except the 388.51 keV gamma radiation from the $^{87}$Rb(p,n)$^{87}$Sr$^m$ reaction - no disturbing gamma lines were observable.

For measuring the induced $\gamma$-activity a lead shielded HPGe detector was used as in our previous (p,n)-study \cite{kiss07}. After each irradiation the $\gamma$ spectra were taken for 12 h. The main experimental challenge was to separate the transition of E$_{\gamma}$ = 514.01 keV from the usually broad annihilation peak coming from beam-induced reactions on impurities of the target and the backing by carefully choosing the waiting and measuring time. 
The 511 keV peak was always less than or comparable to the one of the relevant transition at 514 keV, as shown in the insets of Fig. 1. Because of the relatively long half life of $^{85\mathrm{g}}$Sr ($T_{1/2}$ = 64.84 d) we were able to repeat the activity measurement for each target after approximately 1 month when the intensity of the 511\,keV radiation is substantially reduced. The spectra taken in the repeated activity measurement in the case of the 2.4 and 3.8 MeV irradiations are shown in the lower panels of Fig. 1. The two measurements yielded consistent cross sections proving the proper separation of the 511\,keV and 514\,keV peaks.

\section{Results and conclusions}

Based on the activity measurement one hour after the irradiation and the repeated activity measurement one month later, two separated analysis were done in the case of the $^{85}$Rb(p,n)$^{85}$Sr$^g$ reaction. The derived cross sections agree within 4\%. The final results were calculated from the average weighted by the statistical uncertainty of the two $\gamma$ countings.

The measured astrophysical $S$ factors ($S(E)=\sigma E^{-1} \exp(-2\pi \eta)$, where $\eta$ is the Sommerfeld parameter for taking into account the Coulomb barrier penetration) obtained from the total cross sections leading to the isomeric and ground state of $^{85}$Sr are compared to theoretical predictions obtained with the NON-SMOKER code \cite{nonsmoker} in Fig. 2. 
Two different proton optical potentials were used as input for the calculations: the well know JLM potential \cite{jeu77, lej80} and a modified JLM potential \cite{kiss07}.
As can be seen in Fig. 2, the theoretical energy dependence of the resulting $S$ factor is slightly steeper than the experimental data in the case of the use of the JLM potential, although there is a general agreement in magnitude. In the energy range covered by the measurement, the proton width is smaller than the neutron width (except close to the threshold) and thus uncertainties in the description of the proton width (and proton transmission coefficient) will fully impact the resulting $S$ factor. Figure 2 shows also the prediction resulting from the use of the JLM potential with the imaginary strength increased by 70\%. This modification was introduced in \cite{kiss07}, leading to an
improvement of the reproduction of the $^{70}$Ge(p,$\gamma$)$^{71}$As as well as $^{76}$Ge(p,n)$^{76}$As data (further comparison to available experimental data can be found also there). 

In the case of the $^{85}$Rb(p,n)$^{85}$Sr reaction we find that the energy dependence of the theoretical $S$ factor --- calculated using the modified proton optical potential --- is changed a in such a way as to show perfect agreement with experimental data. This fact supports the conclusions of previous work \cite{kiss07}. 

\begin{figure}
  \rotatebox{270} {\includegraphics[height=.37\textheight]{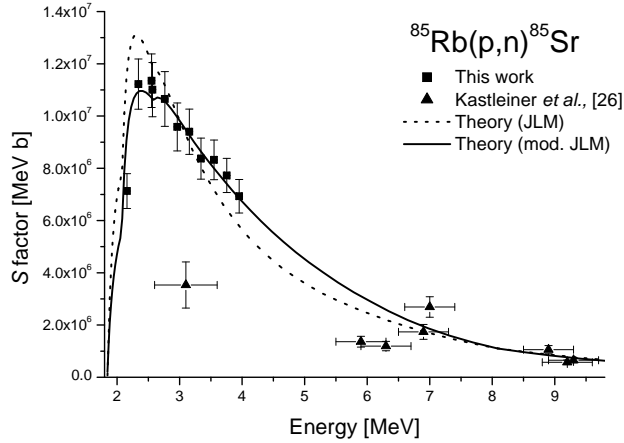}}
  \caption{\label{fig:sfact}Astrophysical $S$ factor of the $^{85}$Rb(p,n)$^{85}$Sr reaction. The lines correspond to Hauser-Feshbach statistical model calculations performed with the NON-SMOKER code \cite{nonsmoker} using different proton optical potentials as input.}
\end{figure}

The $^{85}$Rb(p,n)$^{85}$Sr reaction was already studied by Kastleiner \textsl{et al}., at several proton energies between E$_{c.m.}$ = 3.1 and 70.6 MeV. Unfortunately, the low energy experimental points has large uncertainty in the center of mass energy (as can be seen in Fig. 2.) --- typically between 0.3-0.5 MeV. Consequently, the accuracy is not sufficient to provide a sensitive probe for the statistical model calculations.

\begin{theacknowledgments}

This work was supported by the European Research
Council grant agreement no. 203175, the Economic Competitiveness 
Operative Programme GVOP-3.2.1.-2004-04-0402/3.0., OTKA (K68801, T49245),
and the Swiss NSF (grant 2000-105328).
Gy.\ Gy.\ acknowledges support from the Bolyai grant. 

\end{theacknowledgments}

\end{document}